\documentstyle [12pt]{article}

\def\b{\begin{equation}} \def\e{\end{equation}}
\def\bd{\begin{displaystyle}} \def\ed{\end{displaystyle}}
\def\ba{\begin{array}} \def\ea{\end{array}}

\def\bee{\begin{enumerate}}
\def\eee{\end{enumerate}}

\def\bes{\begin{eqnarray*}}
\def\ees{\end{eqnarray*}}
\def\be{\begin{eqnarray}}
\def\ee{\end{eqnarray}}

\begin{document}
\title{ On the effective action of the vacuum photon splitting in Lorentz violating QED through Krein space}
\author{A. Refaei$^{}$ \thanks{e-mail:
refaei@iausdj.ac.ir}}

\maketitle \centerline{\it Department of Physics,  Sanandaj branch, Islamic Azad
University, Sanandaj, Iran.}

\begin{abstract}
The finding of an explicit effective action describing amplitude of photon triple splitting is investigated through Krein space quantization and quantum metric fluctuation. Using the derivative expansion method, the one-loop radiative corrections from Lorentz- and CPT- violating extended Krein QED shows that it is not possible to find a nonzero effective action.
\end{abstract}

\section{Introduction}
Violation of Lorentz and CPT symmetry in relativistic quantum field theory has not been seen yet. But, if it exists, our understanding about the nature and its symmetry will be disarranged and special relativity will encounter to a serious challenge. However, recently some attempts have been carried out to analyze the several aspects of this possibility, breaking of Lorentz invariance and CPT too \cite{cfj,coko1,coko2,cogl,kospic}. The study of the induction of a photon triple splitting term is an important result in the investigation of the Lorentz symmetry violation\cite{kospic}. In this article, finding the amplitude for vacuum photon splitting in the one-loop radiative corrections from the context of the Lorentz- and CPT- violating QED extension, is investigated by using the derivative expansion of fermion determinants. We will follow the lines and recipes which have been presented in \cite{fapasa}. A nonzero effective action has not been isolated and it has still remained as an open problem. Our method is based on Krein space quantization with quantum metric fluctuation for a spinor field. Through our searching for generating quantum radiatively a nonzero term responsible for the amplitude of splitting, we find the same result for the induction of a photon triplet splitting action.

  Several problems in QFT have been analysed in this manner and some interesting results have been gained \cite{reta1,reta2,foza,fo,paym,payf,re3}. The physical important results, such as one-loop QED effective action, lamb-shift, magnetic anomaly and Euler-Hiesenberg Lagrangian, which have gained in this way, are found in the references \cite{reta1,reta2,foza,re3}. In this research, beside its association with the normal QED, its ability to preserve the gauge invariance is an important problem. Krein space quantization has been appeared as a regularized method and the presence of the negative norm states (proposed by Dirac in 1942 \cite{dir}) although they do not propagate in the physical space, play a renormalizing role \cite{gazeau00,ta3,ta4,rota,gagarota,gaj,sb,derotata,taj}. This quantization method, for the first time, was applied to the minimally coupled scalar field in de Sitter space . It was proved that a covariant quantization of the minimally coupled scalar field in de Sitter space demands addition of the negative frequency solutions of the field equation (or negative norm states) in order to eliminate infrared and
ultraviolet divergences, which maintain causality\cite{gazeau00}.

The organization of this work is as follows:
In section 2, we discuss photon triple splitting and evaluation of its effective action as a theoretical supporting of the discussion. The successor propagator via Krein quantization in metric fluctuated is found in section 3. In section 4, as the final goal of this work we see that a nonzero photon triple splitting effective action can't be isolated.
\section{Photon triple splitting effective action}
We work within the sector of the general QED for a single Dirac field $\psi$ of mass $m$ with an axial vector term
\b {\mathcal{L}}=\overline{\psi}(i\not{\partial}- m -\not{b} \gamma_5-e \not{A})\psi, \e
where $b_{\mu}$ is a constant four-vector with a fixed orientation in space-time \cite{fapasa,kospic,alt}. The term $\overline{\psi}\not{b}\gamma_5\psi$ is gauge invariant and violates Lorentz- and CPT- symmetries explicitly, since $b_{\mu}$ picks up a preferred direction in space-time.
By integrating out the fermion field $\psi$ in the functional integral, the effective action for the gauge field $A_{\mu}(x)$ reads
\b \Gamma[b, A]=-i Tr\ln[\not{p} -m-\not{b}\gamma_5-e\not{A}]. \e
The symbol Tr stands for the trace over Dirac matrices, trace over the internal space as well as for the integrations in momentum and coordinate spaces. One can write
$$ \Gamma[b, A]=\Gamma[b]+\Gamma^{(n)}[b, A], $$
where
\begin{eqnarray}\label{eff1} \Gamma^{(n)}[b, A]&=& -i Tr\ln[1-\frac{e\not{A}}{\not{p} -m-\not{b}\gamma_5}]\nonumber\\
&=&-iTr\sum_{n=1}^{\infty}\frac{1}{n}\left[-ieG^{b}(p)\not{A}(x)\right]^n .\end{eqnarray}
All of the terms in the expansion are one loop contribution, but we focus on the fourth term as the one that will search for an induced photon triple splitting term in the effective action. Then
\b \Gamma^{(4)}[b, A]=-\frac{ie^4}{4}$$$$Tr\Big[G^{b}(p)\not{A}(x)G^{b}(p)\not{A}(x)  G^{b}(p)\not{A}(x)G^{b}(p)\not{A}(x)\Big].\e
By applying properties of derivative expansion, using some techniques of calculation as presented in \cite{fapasa},
in order to keep the full dependence on $b_{\mu}$ it may be rationalized to obtain\cite{fapasa,per},
\begin{eqnarray}
  G^{b}(k)&=&i\frac{(\not{p}-m-\not{b}\gamma_5)(p^2-m^2-b^2+[\not{p},\not{b}]\gamma_5)}{(p^2-m^2-b^2)^2+4(p^2b^2-(p.b)^2)},
\end{eqnarray}
where we use a hermitian $\gamma_5$ with $tr\{\gamma^{\mu}\gamma^{\nu}\gamma^{\rho}\gamma^{\sigma}\gamma_5\}=4i\epsilon ^{\mu\nu\rho\sigma}$ and $g_{\mu\nu}=diag\{1,-1,-1,-1\}$. This form represents one of the ways in which the presence of $b$ will affect the photon splitting effect. By applying of this exact form into the calculation procedure, one expands $G^b$ and take the leading order in $b$ as in the form
 \begin{eqnarray}
  G^{b}(k) &=& G(p)+G(p)(-i\not{b}\gamma_5)G(p)+\cdots,
\end{eqnarray}
where $G(p)=\frac{i}{\not{p}-m}$. So, the trace of gamma matrices to isolate the effective action is
\b \label{eff2} \Gamma_{eff}^{(4)}[b,A]=2\int d^4x\Pi^{\mu\rho\nu\alpha\beta}_{abcd}(\partial_\rho A_\mu)A_\nu A_\alpha A_\beta,
\e
where$$\Pi^{\mu\rho\nu\alpha\beta}_{abcd}=\Pi_{a}^{\mu\rho\nu\alpha\beta}+\Pi_{b}^{\mu\rho\nu\alpha\beta}
+\Pi_{c}^{\mu\rho\nu\alpha\beta}+\Pi_{d}^{\mu\rho\nu\alpha\beta},$$ and $\Pi_{a}^{\mu\rho\nu\alpha\beta},\Pi_{b}^{\mu\rho\nu\alpha\beta}
,\Pi_{c}^{\mu\rho\nu\alpha\beta},\Pi_{d}^{\mu\rho\nu\alpha\beta}$ are self-energy tensors written as below
\begin{eqnarray}
\Pi_{a}^{\mu\rho\nu\alpha\beta}&=& 5ie^4b_\lambda\epsilon^{\mu\rho\nu\lambda}\int \frac{d^4p}{(2\pi)^4}\frac{(p^2-m^2)\eta^{\alpha\beta}-6p^\alpha p^\beta}{(p^2-m^2)^4},\nonumber\\
\Pi_{b}^{\mu\rho\nu\alpha\beta}&=& 6ie^4b_\lambda\epsilon^{\mu\sigma\rho\lambda}\int \frac{d^4p}{(2\pi)^4}\frac{3(p^2-m^2)p_\sigma p^\nu\eta^{\alpha\beta}-8p_\sigma p^\nu p^\alpha p^\beta}{(p^2-m^2)^5},\nonumber\\
\Pi_{c}^{\mu\rho\nu\alpha\beta}&=& 6ie^4b_\lambda\epsilon^{\sigma\nu\rho\lambda}\int \frac{d^4p}{(2\pi)^4}\frac{3(p^2-m^2)p_\sigma p^\mu\eta^{\alpha\beta}-8p_\sigma p^\mu p^\alpha p^\beta}{(p^2-m^2)^5},\nonumber\\
\Pi_{d}^{\mu\rho\nu\alpha\beta}&=& 6ie^4b_\lambda\epsilon^{\mu\mu\sigma\lambda}\int \frac{d^4p}{(2\pi)^4}\frac{3(p^2-m^2)p_\sigma p^\rho\eta^{\alpha \beta}-8p_\alpha p^\rho p^\alpha p^\beta}{(p^2-m^2)^5}.
\label{111}
\end{eqnarray}
It has been shown that these tensors independently vanish. Then, a nonzero photon triple splitting effective action cannot be isolated \cite{fapasa}.

\section{Krein space and Dirac propagator }
As a well-behavior alternative in the quantization methods, we introduce the Krein space quantization. The essential point in this structure is the unavoidable presence of the negative norm states which was proposed by Dirac in 1942 for the first time. They do not propagate in the physical space but they are auxiliary states playing a renormalizing role. Krein space has been constructed by the combination of Hilbert, ${\mathcal{H_{+}}}$, and anti-Hilbert, ${\mathcal{H_{-}}}$, spaces ($\mathcal{H}=\mathcal{H_{+}}\oplus \mathcal{H_{-}}$), where plus and minus indices stand for the positive and negative frequency solutions respectively.

 Let us start with Dirac equation
  \b \label{eq1}
(i\gamma^{\mu}\partial_{\mu}-m)\psi=(i\not{\partial}-m)\psi=0.\e
 Two sets of solutions are given by:
\b u_\alpha(k,x)=\sqrt{\frac{{m}}{{k_0}(2\pi)^3}}u_\alpha(k)e^{-ik.x},\e
\b v_\alpha(k,x)=\sqrt{\frac{{m}}{{k_0}(2\pi)^3}}v_\alpha(k)e^{ik.x},\e
 where $ k^0=(\vec k .\vec k+m^2)^{\frac{1}{2}} \equiv w(\vec k)\geq 0$, note that $v_\alpha$ has the
negative norm.
In Krein QFT we can decompose the spinor field into positive and negative norm parts:
 \b \psi(x)=\frac{1}{\sqrt
2}[\psi^p(x)+\psi^n(x)],\e
where $$ \psi^p(x)=\int d^3\vec k \sum_{\alpha=1,2}\Big[a(\vec k)u_\alpha(k,x)+a^{\dag}(\vec
k)u^{*}_\alpha(k,x)\Big],$$ $$ \psi^n(x)=\int d^3\vec k \sum_{\alpha=1,2}\Big[b(\vec k)v_\alpha(k,x)+b^{\dag}(\vec
k)v^{*}_\alpha(k,x)\Big].$$  $a(\vec k)$ and $b(\vec
k)$ are two independent operators. Creation and annihilation operators are constrained to obey the
following commutation rules
\begin{equation}
[a(\overrightarrow{k}),a(\overrightarrow{k^{'}})]=0,\quad[a^\dagger(\overrightarrow{k}),a^\dagger(\overrightarrow{k^{'}})]=0,
\label{9}
\end{equation}
\begin{equation}
[b(\overrightarrow{k}),b(\overrightarrow{k^{'}})]=0,\quad [b^\dagger(\overrightarrow{k}),b^\dagger(\overrightarrow{k^{'}})]=0,$$$$
[a(\overrightarrow{k}),a^\dagger(\overrightarrow{k^{'}})]=
-[b(\overrightarrow{k}),b^\dagger(\overrightarrow{k^{'}})]=
\delta^3(\overrightarrow{k}-\overrightarrow{k^{'}}),
\label{10}
\end{equation}
\begin{equation}
[a(\vec{k}),b(\overrightarrow{k^{'}})]=0,\quad[a^\dagger(\overrightarrow{k}),b^\dagger(\overrightarrow{k^{'}})]=0,$$$$
[a(\overrightarrow{k}),b^\dagger(\overrightarrow{k^{'}})]=0,\quad [a^\dagger(\overrightarrow{k}),b(\overrightarrow{k^{'}})]=0.
\label{11}
\end{equation}
The vacuum state, $|0>$, is then defined by
$$ a^\dagger(\overrightarrow{k})|0>= |1_{\overrightarrow{k}}>;  \quad   a(\overrightarrow{k})|0>=0  ,\forall \overrightarrow{k}$$
$$ b(\overrightarrow{k})|0>=0;\quad  b^\dagger(\overrightarrow{k})|0>=|\bar{1}_{\overrightarrow{k}}>,
\forall \overrightarrow{k} $$
\begin{equation}
  b(\overrightarrow{k})|1_{\overrightarrow{k}}>=0;
\quad a(\overrightarrow{k})|\bar{1}_{\overrightarrow{k}}>=0,\forall \overrightarrow{k}
\label{14}
\end{equation}
where $| 1_{\vec{k}} >$ is called a one ''particle state'' and $|\bar{1}_{\vec{k}} >$ is called a one ''unparticle state''. These
commutation relations, together with the normalization of the vacuum $$<0|0>=1,$$ lead to positive (resp. negative) norms on the physical (resp. unphysical) sector:
\b
<1_{\overrightarrow{k^{'}}}|1_{\overrightarrow{k}}>
=-<\bar{1}_{\overrightarrow{k^{'}}}|\bar{1}_{\overrightarrow{k}}>=\delta^3(\overrightarrow{k}-\overrightarrow{k^{'}}).\e
The time-ordered product for the spinor field is
\b <S_T(x-x')>\equiv(i\not{\partial}+m)<G_T(x,x')>, \label{sp}\e
where
 \b iS_T(x,x')=<0\mid T\psi(x)\bar{\psi}(x') \mid 0>.\e
In this case we can write
  \b G_T(x,x')=\frac{1}{2}[G_F(x,x')+(G_F(x,x'))^*]=\Re G_F(x,x'),\e
where the Feynman Green function is defined by \cite{bida}
\b G_F(x,x')=\int \frac{d^4 p}{(2\pi)^4}e^{-ip.(x-x') }\tilde
G_F(p)$$$$=\int \frac{d^4p}{(2\pi)^4}\frac{e^{-ip.(x-x')}}{p^2-m^2+i\epsilon} \label{eq:2}
=-\frac{1}{8\pi}\delta
(\sigma_0)$$$$+\frac{m^2}{8\pi}\theta(\sigma_0)\frac{J_1
(\sqrt{2m^2\sigma_0})-iN_1 (\sqrt{2m^2\sigma_0})}{\sqrt{2m^2
\sigma_0}} $$$$-\frac{im^2}{4\pi^2}\theta(-\sigma_0)\frac{K_1
(\sqrt{-2m^2\sigma_0})}{\sqrt{-2m^2 \sigma_0}},\e
 where
$\sigma_0=\frac{1}{2}(x-x')^2 .$ So we have \b G_T(x,x')=\int
\frac{d^4 p}{(2\pi)^4}e^{-ip.(x-x')}{\cal
PP}\frac{1}{p^2-m^2}$$$$ =-\frac{1}{8\pi}\delta
(\sigma_0)+\frac{m^2}{8\pi}\theta(\sigma_0)\frac{J_1
(\sqrt{2m^2\sigma_0})}{\sqrt{2m^2 \sigma_0}},
\label{28}
\e
$x\neq x',$ and ${\cal PP}$ stands for the principal parts. Contribution of the
coincident point singularity $(x=x')$ merely appears in the
imaginary part of $G_F$ (\cite{ta3} and equation (9.52) in \cite{bida})
$$ G_F(x,x)=-\frac{2i}{(4\pi)^2}\frac{m^2}{d-4}+G_F^{\mbox{finit}}(x,x),$$ where $d$ is the space-time dimension and
$G_F^{\mbox{finit}}(x,x)$ becomes finite as $d\longrightarrow 4$.
Note that the singularity of the Eq.(\ref{28}) takes place only on
the cone \emph{i.e.,} $x\neq x', \sigma_0=0$.

It has been shown that the quantum metric fluctuations remove the
singularities of Green's functions on the light-cone \cite{ford97}.
Therefore, the quantum field theory in Krein space, including the
quantum metric fluctuation $\left(
g_{\mu\nu}=\eta_{\mu\nu}+h_{\mu\nu}\right)$, removes all the
ultraviolet divergencies of the theory \cite{rota,ford97}, so one
can write: \begin{eqnarray} \label{a.210}\langle G_T(x,x')\rangle &=& -\frac{1 }{8\pi}
\sqrt{\frac{\pi}{2\langle\sigma_1^2\rangle}}
\exp\left(-\frac{\sigma_0^2}{2\langle\sigma_1^2\rangle}\right)\nonumber\\&+&
 \frac{m^2}{8\pi}\theta(\sigma_0)\frac{J_1(\sqrt {2m^2
 \sigma_0})}{\sqrt {2m^2 \sigma_0}},\end{eqnarray}
where $2\sigma= g_{\mu\nu}(x^{\mu}-
x'^{\mu})(x^{\nu} - x'^{\nu})$ and $\sigma_1$ is the first order shift in $\sigma$, due to
the linear quantum gravity ($\sigma=\sigma_0+\sigma_1+\emph{O}(h^2)$). The average value is taken over the quantum metric fluctuation and in the case of $2\sigma_0 =\eta_{\mu\nu}(x^{\mu}- x'^{\mu})(x^{\nu} - x'^{\nu})=0$ we have
$\langle\sigma_1^2\rangle\neq 0$. So, we get \b \langle
G_T(0)\rangle = -\frac{1 }{8\pi}
\sqrt{\frac{\pi}{2\langle\sigma_1^2\rangle}} +
 \frac{m^2}{8\pi}\frac{1}{2}.\e
It should be noted that $ \langle\sigma_1^2\rangle $ is related to
the density of gravitons \cite{ford97}.

By using the Fourier transformation of Dirac delta function,
$$ -\frac{1}{8\pi}\delta(\sigma_0)= \int \frac{d^4
p}{(2\pi)^4}e^{-ip.(x-x')}{\cal PP}\frac{1}{p^2},$$or equivalently
$$\frac{1}{8\pi^2}\frac{1}{\sigma_0}= -\int \frac{d^4
p}{(2\pi)^4}e^{-ip.(x-x')}\pi\delta(p^2),$$ for the second part of
Green function in (\ref{a.210}), we obtain \b
\int \frac{d^4
p}{(2\pi)^4}e^{-ip.(x-x')}{\cal PP}\frac{m^2}{p^2(p^2-m^2)}\e
and for the first part we have $$
\int \frac{d^4
p}{(2\pi)^4}e^{-ik.(x-x') }  \tilde{G}_1(p),$$
 where $\tilde{G_1}$ is fourier transformation of the first part of the Green function (\ref{a.210}). Therefore, we
obtain \b \label{eq:22}<\tilde G_T(p)>=\tilde{G}_1(p)+{\cal
PP}\frac{m^2}{p^2(p^2-m^2)}.\e It has been proved
that in the one-loop approximation, the Green function in Krein
space quantization, which appears in the transition amplitude is: \begin{eqnarray} <\tilde G_T(p)>\mid_{\mbox{one-loop}}&\equiv & \tilde
G_T(p)\mid_{\mbox{one-loop}}\nonumber\\&\equiv &{\cal PP}
\frac{m^2}{p^2(p^2-m^2)}.
\label{213} \end{eqnarray}
 That means in the one-loop
approximation, the contribution of $\tilde{G_1}(p)$ is negligible.
It is worth mentioning that in order to improve the UV behavior in
relativistic higher-derivative correction theories, the propagator
(\ref{213}) has been used by some authors \cite{bach,ho,ka}.

Therefor, from equations (\ref{sp}) and (\ref{213}), the Fourier transformation of spinor propagator becomes
\begin{eqnarray} (\not{p}+m)\Big[{\cal PP}
\frac{m^2}{p^2(p^2-m^2)}\Big].
\end{eqnarray}
\section{ Effective action in Krein space}
By preparing the new propagator, we are ready to evaluate the photon self-energy tensors in our new approach. In the equations (2.8), $\frac{1}{p^2-m^2}$ term can be replaced by, $\frac{m^2}{p^2(p^2-m^2)},$ as a Green function in Krein space quantization with metric fluctuation (${\mathcal{PP}}$ symbol has been vanished because the $\epsilon$ value has been equaled to zero). Consequently, the self-energy tensors in the new approach and in a compact form become
  \begin{eqnarray}
\Pi_{K,a}^{\mu\rho\nu\alpha\beta}&=& 5im^6e^4b_\lambda\epsilon^{\mu\rho\nu\lambda}\left(\eta^{\alpha\beta}I_3-6m^2I^{\alpha \beta}_4\right),
\nonumber\\
\Pi_{K,b}^{\mu\rho\nu\alpha\beta}&=&6im^8e^4b_\lambda\epsilon^{\mu\sigma\rho\lambda}\left(3\eta_{\theta\sigma} \eta^{\alpha\beta} I^{\theta\nu}_4-8m^2\eta_{\theta\sigma}I^{\theta\nu \alpha \beta}_5\right),
\nonumber\\
\Pi_{K,c}^{\mu\rho\nu\alpha\beta}&=& 6im^8e^4b_\lambda\epsilon^{\sigma\nu\rho\lambda }\left(3\eta_{\theta\sigma} \eta^{\alpha\beta}I^{\theta \mu}_4-8m^2\eta_{\theta\sigma}I^{\theta\mu\alpha\beta}_5\right),
\nonumber\\
\Pi_{K,d}^{\mu\rho\nu\alpha\beta}&=& 6im^8e^4b_\lambda\epsilon^{\mu\nu\sigma\lambda}\left(3\eta_{\theta\sigma}\eta^{\alpha\beta}I^{\theta \rho}_4-8m^2\eta_{\theta\sigma}I^{\theta \rho \alpha \beta}_5\right),
\label{tensap1}
 \end{eqnarray}
where we have
 \begin{eqnarray} I_3&=& \int \frac{d^4p}{(2\pi)^4}\frac{1}{[p^2(p^2-m^2)]^3},\nonumber\\  I_4^{\lambda\nu}&=&\int \frac{d^4p}{(2\pi)^4}\frac{p^\lambda p^\nu}{[p^2(p^2-m^2)]^4},\nonumber\\  I_5^{\sigma\mu\alpha\beta}&=&
\int \frac{d^4p}{(2\pi)^4}\frac{p^\sigma p^\nu p^\alpha p^\beta}{[p^2(p^2-m^2)]^5}.\label{is}\end{eqnarray}
Now, we may calculate the above integrals explicitly. The argument of denominators may be changed to a complete square and in 4-D Minkowski space by Wick rotation, $p_0\rightarrow ip_0$, we can write $d^4p=i2\pi^2p^3dp.$ Also, one can take $k=p^2+\frac{m^2}{2}$ and $dk=2pdp.$ The expressions in (\ref{is}) are decomposed as

 \begin{eqnarray} I_3&=& \int i\frac{2\pi^2p^3dp}{(2\pi)^4}\frac{1}{[(p^2+\frac{m^2}{2})^2-\frac{m^4}{4}]^3}\nonumber\\&=&\int i\frac{(k-\frac{m^2}{2})dk}{16\pi^2}\frac{1}{(k^2-\frac{m^4}{4})^3}\nonumber\\&=& \frac{i}{16\pi^2}\int \frac{kdk}{(k^2-\frac{m^4}{4})^3}- \frac{im^2}{32\pi^2}\int \frac{dk}{(k^2-\frac{m^4}{4})^3},\label{i1}\end{eqnarray}
 in the numerator of the symmetric integrand we replaced $p^\lambda p^\nu$ by $\frac{1}{4}\eta^{\lambda\nu}p^2,$ then
\begin{eqnarray}I^{\lambda\nu}_4&=& \int i\frac{2\pi^2p^3dp}{(2\pi)^4}\frac{-\frac{1}{4}\eta^{\lambda\nu}p^2}{[(p^2+\frac{m^2}{2})^2-\frac{m^4}{4}]^4} \nonumber\\ &=& -\frac{i}{4}\int \frac{(k-\frac{m^2}{2})dk}{16\pi^2}\frac{\eta^{\lambda\nu}(k-\frac{m^2}{2})}{(k^2-\frac{m^4}{4})^4}\nonumber\\
&=& -\frac{1}{4}\frac{i\eta^{\lambda\nu}}{16\pi^2}\Big(\int \frac{k^2dk}{(k^2-\frac{m^4}{4})^4}-m^2\int \frac{kdk}{(k^2-\frac{m^4}{4})^4}\nonumber\\&+&\frac{m^4}{4}\int \frac{dk}{(k^2-\frac{m^4}{4})^4} \Big),\label{i2}\end{eqnarray}
also, in $I^{\sigma\nu\alpha\beta}_n$, we will substitute $p^\sigma p^\nu p^\alpha p^\beta=\frac{1}{24}G^{\sigma\nu\alpha\beta}p^4$ and after all the changes we have
\begin{eqnarray}\label{i3} I^{\sigma\nu\alpha\beta}_5&=&\int \frac{d^4p}{(2\pi)^4}\frac{p^\sigma p^\nu p^\alpha p^\beta}{[p^2(p^2-m^2)]^5}\nonumber\\&=& \int i\frac{2\pi^2p^3dp}{(2\pi)^4}\frac{\frac{1}{24}G^{\sigma\nu\alpha\beta}p^4}{[(p^2+\frac{m^2}{2})^2-\frac{m^4}{4}]^5}\nonumber\\ &=& \int i \frac{dk}{16\times 24\pi^2}\frac{G^{\sigma\nu\alpha\beta}(k-\frac{m^2}{2})^3}{(k^2-\frac{m^4}{4})^5}\nonumber\\ &=&\frac{iG^{\sigma\nu\alpha\beta}}
{16\times 24\pi^2}\Big(\int \frac{k^3dk}{(k^2-\frac{m^4}{4})^5}-\frac{3m^2}{2}\int \frac{k^2dk}{(k^2-\frac{m^4}{4})^5}\nonumber\\ &+&\frac{3m^4}{4}\int \frac{kdk}{(k^2-\frac{m^4}{4})^5}-\frac{m^6}{8}\int \frac{dk}{(k^2-\frac{m^4}{4})^5} \Big).\end{eqnarray}
where $G^{\sigma\nu\alpha\beta}=\eta^{\sigma\nu}\eta^{\alpha\beta}+\eta^{\sigma\alpha}\eta^{\nu\beta}+\eta^{\sigma\beta}\eta^{\nu\alpha}.$ Evaluating of the equations (\ref{i1}), (\ref{i2}) and (\ref{i3}) have a lot of details, hence, one can rewrite these expressions in a compact form as
 \begin{eqnarray}\label{a.14} I_3&=& \frac{i}{16\pi^2}\Big(J^1_3- \frac{m^2}{2}J^0_3\Big).\end{eqnarray}
\begin{eqnarray}\label{a.15}I^{\lambda\nu}_4
 &=&-\frac{1}{4}\frac{i\eta^{\lambda\nu}}
{16\pi^2}\Big(J^2_4-m^2J^1_4 +\frac{m^4}{4}J^0_4 \Big).\end{eqnarray}
\begin{eqnarray}\label{a.16}I^{\sigma\nu\alpha\beta}_5&=&\frac{1}{24}\frac{iG^{\sigma\nu\alpha\beta}}
{16\pi^2}\Big(J^3_5-\frac{3m^2}{2}J^2_5+\frac{3m^4}{4}J^1_5-\frac{m^6}{8}J^0_5 \Big).\end{eqnarray}
where
\b J^j_{n}=\int \frac{k^j dk}{(k^2-\frac{m^4}{4})^n}, \quad \begin{array}{c}
        n=3  \\
         j=0,1
       \end{array},\quad \begin{array}{c}
        n=4  \\
         j=0,1,2
       \end{array},\quad \begin{array}{c}
        n=5 \\
         j=0,1,2,3
       \end{array}.\e
In appendix we have taken solutions of $J^j_{n}$ in different values of j, and $I_3, I^{\lambda\nu}_4$ and $I^{\sigma\nu\alpha\beta}_5$ expressions have obtained as
\b I_3=\frac{3i}{16\pi^2m^8}\ln\frac{m^2}{\mu^2},\quad I^{\lambda\nu}_4=-\frac{i\eta^{\lambda\nu}}{16\pi^2m^{10}}\ln\frac{m^2}{\mu^2}
,\quad I^{\sigma\nu\alpha\beta}_5=\frac{5i G^{\sigma\nu\alpha\beta}}{16\pi^2 \times 24 m^{10}}\ln\frac{m^2}{\mu^2} \label{i},\e
where $\mu^2$ is the scale of energy and tends to zero. We notice that, by using the items of equations (\ref{i}) into equations (\ref{tensap1}), the self-energy tensors become
\b
\Pi_{K,a}^{\mu\rho\nu\alpha\beta}=-\frac{45e^4b_\lambda\epsilon^{\mu\rho\nu\lambda}\eta^{\alpha\beta}}{16\pi^2m^2}\ln\frac{m^2}{\mu^2},\quad
\Pi_{K,b}^{\mu\rho\nu\alpha\beta}=\Pi_{K,c}^{\mu\rho\nu\alpha\beta}=\Pi_{K,d}^{\mu\rho\nu\alpha\beta}= -\frac{9e^4b_\lambda\epsilon^{\mu\rho\nu\lambda}\eta^{\alpha\beta}}{\pi^2m^2}\ln\frac{m^2}{\mu^2},
\label{tensap2}
 \e
 and the sum all of the tensors is
 \b
\Pi_{K,a,b,c,d}^{\mu\rho\nu\alpha\beta}=-\frac{189e^4b_\lambda\epsilon^{\mu\rho\nu\lambda}\eta^{\alpha\beta}}{16\pi^2}\frac{1}{m^2}\ln\frac{m^2}{\mu^2}.
\label{tensap3}\e
 At low energy, $p^2=0$, and to avoid the gauge invariant breaking and the lack of divergency in the theory, it should be applied the large $m$ limit or equivalently, $m^2 \rightarrow \infty.$ It is clear that
 \b
\lim_{\begin{array}{c}
         \mu^2\rightarrow 0, \\
         m^2\rightarrow \infty
       \end{array}}\frac{1}{m^2}\ln\frac{m^2}{\mu^2}=0,
\label{tensap0}
 \e
    which leads to
  \b
\Pi_{K,a,b,c,d}^{\mu\rho\nu\alpha\beta}=0.
 \e

Then a nonzero photon triple splitting effective action can not be isolated which is in line with the results in \cite{kospic} and \cite{fapasa}. One may find the Krein regularization in quantum metric fluctuated as a reliable way in obtaining physical results. As we know, the negative norm states and quantum fluctuation of the metric especially play the main role in this research and the outstanding advantage of the method is the preservation of gauge invariance.

\section{Conclusion}
In summary, we have discussed the Lorentz- and CPT violating extended QED to investigate the induction of an amplitude for vacuum photon splitting by using the derivative expansion of fermion determinants through Krein regularization. An effective action describing amplitude of photon triplet splitting can not be found in the presented method. One of the noticeable point of this method is its flexibility in the preserving of the gauge invariance. Our result is the same as the one which was investigated in the usual QED. The combination of the negative norm states and metric fluctuated in QFT, provides not only a regularized theory but also its consequences coincide with the normal theory results, at least in this study.
\vskip 0.5 cm

\noindent {\bf{Acknowledgments}}: I would like to thank Prof. M. V. Takook and Prof. F. Kheirandish for very useful discussion.

\begin{appendix}
\setcounter{equation}{0}
\section{Some details of calculations}
Here we give the solutions of $J^i_n$ in (\ref{a.14}), (\ref{a.15}) and (\ref{a.16}) equations and finally we find the expressions of $I_3$, $I^{\lambda\nu}_4$ and $I^{\sigma\nu\alpha\beta}_5$
\begin{eqnarray}  \nonumber J^0_3 &=& \int \frac{dk}{(k^2-\frac{m^4}{4})^3}=
\frac{96k^3}{m^8\big(4k^2-m^4\big)^2}-\frac{40k}{m^4\big(4k^2-m^4\big)^2}
-\frac{6}{m^{10}}\ln\frac{k+\frac{m}{2}}{k-\frac{m}{2}}\\ \Rightarrow
\lim_{p^2\rightarrow 0}J^0_3&=& -\frac{1}{2m^6p^4}-\frac{6}{m^{10}}\ln(1+\frac{m^2}{p^2}).\\  J^1_3&=&\int \frac{kdk}{(k^2-\frac{m^4}{4})^3}=-\frac{4}{\big(-4k^2+m^4\big)^2}\Rightarrow\lim_{p^2\rightarrow 0}J^1_3=-\frac{1}{4m^{4}p^4}.
 \\ I_3 &=& \frac{i}{16\pi^2}\lim_{p^2\rightarrow 0}\big(J^1_3-\frac{m^2}{2}J^0_3\big)=\frac{3i}{16\pi^2m^8}\ln\frac{m^2}{p^2}.\end{eqnarray}

\begin{eqnarray}\nonumber J^0_4&=&\int \frac{dk}{(k^2-\frac{m^4}{4})^4}=\frac{8\times 160 k^5}{m^{12}\big(-4k^2+m^4\big)^3}-\frac{8\times320 k^3}{3m^{8}\big(-4k^2+m^4\big)^3}\\ \nonumber&+&\frac{8\times 22 k}{m^{4}\big(-4k^2+m^4\big)^3}+\frac{20}{m^{14}}\ln\frac{k+\frac{m}{2}}{k-\frac{m}{2}} \\ \Rightarrow\lim_{p^2\rightarrow 0}J^0_4&=&-\frac{1}{3m^{8}p^6}+\frac{20}{m^{14}}\ln(1+\frac{m^2}{p^2}).\\ J^1_4&=&\int \frac{kdk}{(k^2-\frac{m^4}{4})^4}=-\frac{1}{6\big(k^2-\frac{m^4}{4}\big)^3}\Rightarrow\lim_{p^2\rightarrow 0}J^1_4=-\frac{1}{6m^{6}p^6}.\\ \nonumber J^2_4&=&\int \frac{k^2dk}{(k^2-\frac{m^4}{4})^4}=-\frac{64 k^5}{m^{8}\big(-4k^2+m^4\big)^3}+\frac{128 k^3}{3m^{4}\big(-4k^2+m^4\big)^3}\\ \nonumber &+& \frac{4 k}{\big(-4k^2+m^4\big)^3}
-\frac{1}{m^{10}}\ln\frac{k+\frac{m}{2}}{k-\frac{m}{2}}\\
\Rightarrow \lim_{p^2\rightarrow 0}J^2_4&=&-\frac{1}{12m^{4}p^6}-\frac{1}{m^{10}}\ln(1+\frac{m^2}{p^2}).\\ I^{\lambda\nu}_4 &=&-\frac{i\eta^{\lambda\nu}}{64\pi^2}\lim_{p^2\rightarrow 0}\big(J^2_4-m^2J^1_4+\frac{m^4}{4}J^0_4\big)=-\frac{i\eta^{\lambda\nu}}{16\pi^2m^{10}}\ln\frac{m^2}{p^2}.\end{eqnarray}

\begin{eqnarray} \nonumber J_5^0 &=&\int \frac{dk}{(k^2-\frac{m^4}{4})^5}=\frac{8}{3}\Big(\frac{ 6720 k^7}{m^{16}(-4k^2+m^4)^4}-\frac{6160 k^5}{m^{12}(-4k^2+m^4)^4}\\ \nonumber &+&\frac{2044 k^3}{m^{8}(-4k^2+m^4)^4}-\frac{279 k}{m^{4}(-4k^2+m^4)^4}\Big)-\frac{70}{m^{18}}\ln(1+\frac{m^2}{p^2})\\ \Rightarrow\lim_{p^2\rightarrow 0}J^0_5&=&-\frac{1}{4m^{10}p^8}-\frac{70}{m^{18}}\ln(1+\frac{m^2}{p^2}).\\ J_5^1&=&-\frac{32}{(-4k^2+m^4)^4}\Rightarrow\lim_{p^2\rightarrow 0}J^1_5=-\frac{1}{8m^{8}p^8}. \\ \nonumber J_5^2&=&\int \frac{k^2dk}{(k^2-\frac{m^4}{4})^5}=\frac{2}{3}\Big(-\frac{960 k^7}{m^{12}(-4k^2+m^4)^4}+\frac{880 k^5}{m^{8}(-4k^2+m^4)^4}\\ \nonumber
&-&\frac{292 k^3}{m^{4}(-4k^2+m^4)^4}+\frac{15 k}{(-4k^2+m^4)^4}\Big)+\frac{5}{2m^{14}}\ln(1+\frac{m^2}{p^2})\\ \Rightarrow\lim_{p^2\rightarrow 0}J^2_5&=&-\frac{1}{8m^{6}p^8}+\frac{5}{2m^{14}}\ln(1+\frac{m^2}{p^2}).\\ J_5^3&=&\frac{8(-16k^2+m^4)}{3(-4k^2+m^4)^4} \Rightarrow\lim_{p^2\rightarrow 0}J^3_5=-\frac{1}{8m^{4}p^8}.\\ \nonumber I^{\sigma\nu\alpha\beta}_5&=&\frac{1}{24}\frac{iG^{\sigma\nu\alpha\beta}}
{16\pi^2}\lim_{p^2\rightarrow 0}\Big(J^3_5-\frac{3m^2}{2}J^2_5+\frac{3m^4}{4}J^1_5-\frac{m^6}{8}J^0_5 \Big)\\
I^{\sigma\nu\alpha\beta}_5&=&\frac{5i G^{\sigma\nu\alpha\beta}}{16\pi^2 \times 24 m^{10}}\ln\frac{m^2}{p^2}.\end{eqnarray}

\end{appendix}

\end{document}